\documentclass[conference]{IEEEtran}
\IEEEoverridecommandlockouts
\usepackage{placeins}
\usepackage{subcaption}
\usepackage{cite}
\usepackage{amsmath,amssymb,amsfonts}
\usepackage{algorithmic}
\usepackage{graphicx}
\usepackage{textcomp}
\usepackage{xcolor}
\usepackage{booktabs}
\usepackage{multirow}
\usepackage{array}
\usepackage{adjustbox}
\usepackage{comment}
\usepackage{pifont}
\usepackage{adjustbox}
\usepackage{pgfplots}
\pgfplotsset{compat=1.18}
\usepgfplotslibrary{polar}

\usetikzlibrary{shapes,arrows,positioning,fit,backgrounds,shapes.geometric,shapes.symbols,calc}

\begin{document}

\title{Device-Native Autonomous Agents for Privacy-Preserving Negotiations\\
}

\author{\IEEEauthorblockN{
1\textsuperscript{st} Joyjit Roy\,
}
\IEEEauthorblockA{\textit{
IEEE Senior Member} \\
Austin, Texas, USA \\
joyjit.roy.tech@gmail.com}
\and
\IEEEauthorblockN{
2\textsuperscript{nd} Samaresh Kumar Singh\, 
}
\IEEEauthorblockA{\textit{
IEEE Senior Member} \\
Leander, Texas, USA \\
ssam3003@gmail.com}
}

\maketitle

\begin{abstract}
Automated negotiations in insurance and business-to-business (B2B) commerce encounter substantial challenges. Current systems force a trade-off between convenience and privacy by routing sensitive financial data through centralized servers, increasing security risks, and diminishing user trust. This study introduces a device-native autonomous Agentic AI system for privacy-preserving negotiations. The proposed system operates exclusively on user hardware, enabling real-time bargaining while maintaining sensitive constraints locally. It integrates zero-knowledge proofs to ensure privacy and employs distilled world models to support advanced on-device reasoning. The architecture incorporates six technical components within an Agentic AI workflow. Agents autonomously plan negotiation strategies, conduct secure multi-party bargaining, and generate cryptographic audit trails without exposing user data to external servers. The system is evaluated in insurance and B2B procurement scenarios across diverse device configurations. Results show an average success rate of 87\%, a 2.4$\times$ reduction in latency relative to cloud baselines, and strong privacy preservation through zero-knowledge proofs. User studies show 27\% higher trust scores when decision trails are available. These findings establish a foundation for trustworthy autonomous agents in privacy-sensitive financial domains.
\end{abstract}


\section{Introduction}
\label{sec:introduction}

Insurance and business-to-business (B2B) commerce typically rely on static pricing models, offering customers fixed quotes that do not allow for negotiation. For example, health insurance applicants cannot trade higher deductibles for lower premiums, and procurement managers cannot negotiate bulk discounts in real time. Although manual bargaining is possible, it remains slow, inconsistent, and limited to high-value transactions, leading to rigid, often suboptimal pricing decisions.

Autonomous agents present a promising direction to address these limitations~\cite{yao2023react,autogpt2023}. In bilateral negotiations, buyer and seller agents conduct real-time bargaining on behalf of their principals. Buyers can specify constraints, such as a maximum budget and acceptable trade-offs, while the agents manage the negotiation process automatically. However, several significant challenges persist: privacy (users must not disclose financial constraints to external servers), explainability (regulatory requirements mandate decision transparency~\cite{goodman2017european}), resource limitations (restricting on-device algorithm complexity), and fairness (preventing exploitative agreements~\cite{nash1950bargaining}).

This work introduces a device-native Agentic AI system for autonomous negotiations. The proposed approach converts static pricing into privacy-preserving bargaining through a six-layer architecture that addresses these challenges through (1) Selective state transfer, (2) Explainable memory, (3) World model distillation ~\cite{hinton2015distilling}, (4) Privacy-preserving protocols with zero-knowledge proofs ~\cite{groth2016size}, (5) Model-aware offloading, and (6) Simulation-critic safety mechanisms.

\section{Related Work}
\label{sec:related}

\subsection{Agentic AI Frameworks}

Recent frameworks have advanced autonomous AI reasoning. ReAct~\cite{yao2023react} integrates reasoning and acting within language models. AutoGPT~\cite{autogpt2023} sequences large language model (LLM) calls to address complex tasks. LangChain~\cite{langchain2023} provides tools for constructing agent applications. Although these systems demonstrate autonomous capabilities, they share limitations. All depend on cloud infrastructure for execution and lack privacy-preserving mechanisms. None supports bilateral negotiation protocols or runs on resource-constrained devices.

\subsection{Multi-Agent Systems and Negotiation}
Game theory provides foundational principles for automated negotiation. Nash bargaining~\cite{nash1950bargaining} defines fair outcomes in bilateral settings. Rubinstein's alternating offers model~\cite{rubinstein1982perfect} describes sequential bargaining dynamics. These theoretical frameworks inform the protocol design presented in this work.

Recent multi-agent systems demonstrate potential for agent cooperation. MetaGPT~\cite{hong2023metagpt} assigns distinct roles to collaborating agents, while CAMEL~\cite{li2023camel} facilitates communication through role-playing mechanisms. However, these systems primarily address collaborative task completion rather than bilateral negotiation involving competing interests. Privacy considerations are minimally addressed in their designs.

Zero-knowledge proofs~\cite{goldwasser1989knowledge} enable private computation without disclosing data. zk-SNARKs~\cite{groth2016size} provide efficient proof generation. This work integrates them into negotiation protocols to demonstrate constraint satisfaction.

Privacy attacks on machine learning systems extend beyond cryptographic leakage. Timing attacks exploit computational patterns to infer sensitive information~\cite{jin2024timing}. Model inversion and membership inference attacks can reconstruct private inputs. While zero-knowledge proofs mitigate direct data exposure, behavioral inference remains a challenge (Sec.~\ref{sec:limitations}).

\subsection{Explainability and Trust}
Research in explainable AI demonstrates that audit trails improve user trust~\cite{ribeiro2016should}. LIME and SHAP~\cite{lundberg2017unified} provide post-hoc explanations for model predictions. Financial AI systems must comply with regulations that require decision transparency~\cite{goodman2017european}.

Current agent systems lack cryptographic auditability, as decision logs typically consist of informal text traces without verifiability guarantees. The proposed explainable memory system employs Merkle trees for tamper-evident logging and utilizes blockchain anchoring to enable independent verification.

Table~\ref{tab:comparison} summarizes the comparison. The proposed framework combines autonomous reasoning, privacy-preserving negotiation, and cryptographic explainability within a device-native architecture. 

\begin{table}[t]
\vspace{0.05in}
\centering
\caption{Comparison of Agentic AI Systems}
\label{tab:comparison}
\adjustbox{max width=\columnwidth}{%
\begin{tabular}{lccccccc}
\toprule
\multirow{2}{*}{System} & \multicolumn{7}{c}{Capabilities} \\
\cmidrule(lr){2-8}
& Auto. & Priv. & Neg. & Expl. & Attest. & Device & State \\
\midrule
AutoGPT~\cite{autogpt2023} & \checkmark & -- & -- & -- & -- & -- & -- \\
LangChain~\cite{langchain2023} & \checkmark & -- & -- & $\sim$ & -- & -- & -- \\
MetaGPT~\cite{hong2023metagpt} & \checkmark & -- & -- & $\sim$ & -- & -- & -- \\
CAMEL~\cite{li2023camel} & \checkmark & -- & -- & -- & -- & -- & -- \\
\midrule
\textbf{Proposed Model} & \checkmark & \checkmark & \checkmark & \checkmark & \checkmark & \checkmark & \checkmark \\
\bottomrule
\end{tabular}%
}
\begin{flushleft}
\footnotesize 
Auto.=Autonomous Reasoning, Priv.=Privacy Preservation (ZK Proofs), \\
Neg.=Bilateral Negotiation, Expl.=Cryptographic Explainability, \\
Attest.=Code Attestation, Device=On-Device Execution, \\
State=Cross-Device State Transfer. \\
\checkmark=Full support, $\sim$=Partial support, --=Not supported.
\end{flushleft}
\end{table}
\section{System Architecture}
\label{sec:architecture}

This section presents the device-native Agentic AI architecture for autonomous negotiations. The workflow is described first, followed by detailed explanations of the 6 components.

\subsection{Agentic Workflow Overview}
\begin{figure*}[htbp]
\centering
\includegraphics[width=0.9\textwidth]{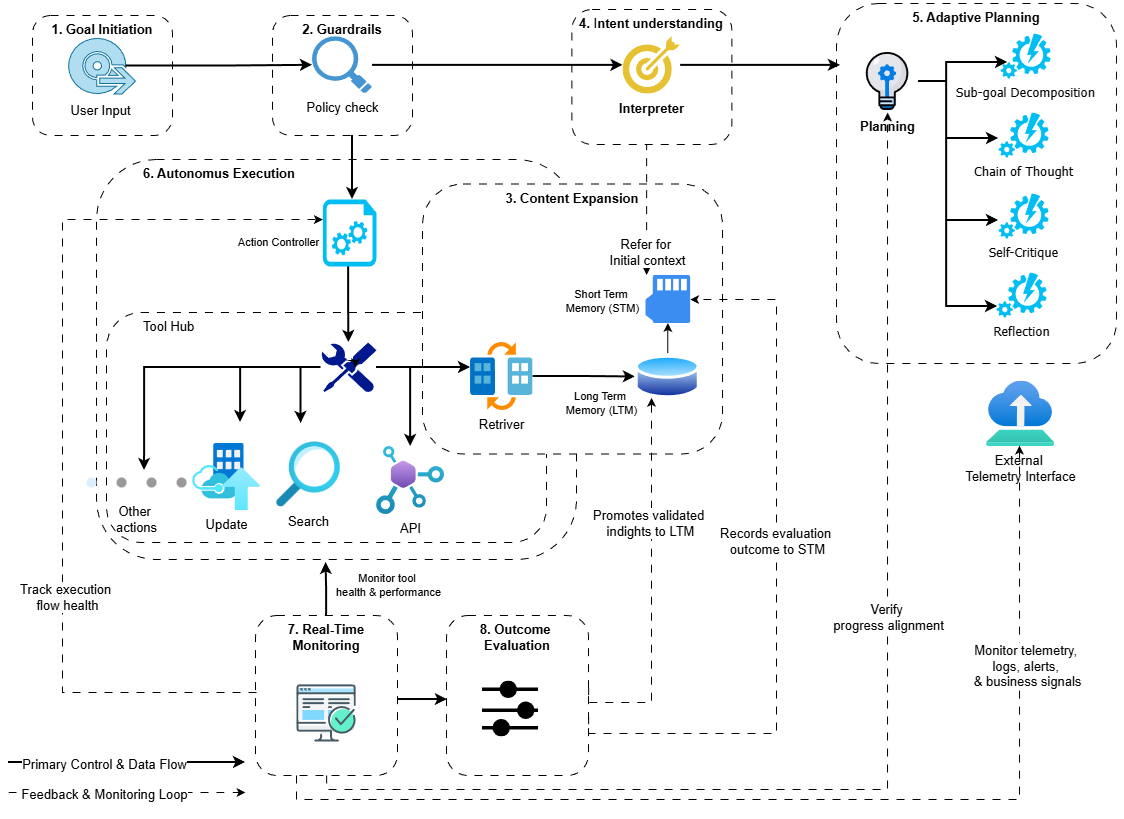}
\caption{Agentic AI workflow architecture showing 8 sequential stages from Goal Initiation to Outcome Evaluation, with Tool Hub and dual-memory (STM/LTM) components.}
\label{fig:workflow}
\end{figure*}

Figure~\ref{fig:workflow} illustrates the eight-step Agentic AI workflow that governs negotiation behavior. The process begins with users defining negotiation objectives in the \textbf{Goal Initiation}, where target price ranges and acceptable tradeoffs are specified. In the subsequent \textbf{Guardrails} phase, these goals are validated against policy constraints to ensure adherence to budget limits and regulatory compliance.

\textbf{Context Expansion} retrieves information from short-term memory (STM), such as the current negotiation state, and long-term memory (LTM), including past transactions, preferences, and domain knowledge. \textbf{Intent Understanding} interprets the negotiation goal within this context, identifies the negotiation type, and anticipates the counterparty's expectations.

During \textbf{Adaptive Planning}, the overarching negotiation goal is decomposed into actionable sub-goals, and a multi-stage negotiation strategy is formulated. The planner assesses strategic approaches, such as aggressive opening offers, incremental concessions, and package deals that combine multiple terms. In the \textbf{Autonomous Execution }phase, the selected strategy is implemented by the action controller, which exchanges offers with counterparty agents through the Tool Hub.

\textbf{Real-Time Monitoring} tracks progress against the planned strategy and initiates replanning when counterparty behavior deviates. \textbf{Outcome Evaluation} assesses the final agreement, updates memory with lessons learned, and generates feedback for future negotiations.

\subsection{Component 1: Selective State Transfer}

Negotiations often extend across multiple sessions and devices. For example, a negotiation may begin on a laptop, continue on a mobile device during transit, and conclude on a tablet at home. The complete negotiation state encompasses conversation history, active offers, counterparty models, and planning context. In complex multi-round negotiations, the state size can reach 8MB.

Transferring 8MB of data over mobile networks introduces significant latency. The selective state transfer mechanism addresses this issue by compressing the negotiation state to 2–4MB using three techniques. First, \textbf{critical state identification} distinguishes essential elements from reconstructable data. Recent messages and active offers are considered critical, while older conversation turns are summarized. Next, \textbf{embedding pruning} reduces redundancy by retaining only the most recent 50 embeddings and clustering older ones. Finally, \textbf{delta encoding} transmits only the changes since the last synchronization point.

The compression algorithm is formalized as follows. Let $S = \{s_1, s_2, \ldots, s_n\}$ represent the full state. Importance scores $I(s_i)$ are calculated based on recency and relevance to the ongoing negotiation. States with scores below the threshold $\tau$ are pruned or summarized:
\begin{equation}
S_{\text{compressed}} = \{s_i : I(s_i) > \tau\} \cup \text{summarize}(\{s_i : I(s_i) \leq \tau\})
\end{equation}

This mechanism is integrated into the \textbf{Context Expansion} phase of the workflow (Figure~\ref{fig:workflow}). When an agent resumes activity on a new device, the compressed state restores the negotiation context. STM receives the active state, while LTM receives the summarized history. Performance results are presented in Table~\ref{tab:state_transfer}.

\subsection{Component 2: Explainable Memory}

Regulated domains require decision audit trails to ensure compliance and accountability. For example, insurance negotiations must document the rationale for accepting or rejecting specific terms, while B2B procurement systems require comprehensive records to facilitate compliance audits. Additionally, transparency helps users understand and trust agent behavior.

The explainable memory system produces cryptographic audit trails for each decision. Each record contains timestamp, decision type, inputs, reasoning trace, and outcome. These records are linked using Merkle trees~\cite{merkle1987digital}, and the root hash is periodically anchored to a public blockchain to enable independent verification.

The memory structure supports three query types. \textbf{Point queries} retrieve specific decisions by identifier or timestamp. \textbf{Range queries} return all decisions within a specified time window. \textbf{Proof queries} generate cryptographic proofs that demonstrate a decision existed at a claimed time with specified content.

Formally, for a decision $d_i$ with content $c_i$, the hash is computed as follows:
\begin{equation}
h_i = H(c_i \,||\, h_{i-1})
\end{equation}
Here, $H$ denotes a collision-resistant hash function, and $h_{i-1}$ represents the hash of the previous decision. Merkle proofs enable verification in $O(\log n)$ time.

This mechanism integrates with \textbf{LTM} and \textbf{Outcome Evaluation} within the workflow (Figure~\ref{fig:workflow}). Decisions are stored in LTM for persistence, and Outcome Evaluation logs final results along with complete reasoning traces. Performance results are presented in Table~\ref{tab:memory_overhead}.

\subsection{Component 3: World Model Distillation}

Effective negotiation requires reasoning about counterparty behavior. Agents must anticipate offer acceptance and response strategies. Planning requires simulating multiple approaches before committing.

Cloud-based reasoning models provide this capability but fail to meet privacy requirements. Since a 7B-parameter model cannot execute on consumer devices, large cloud models are distilled into compact 500M-parameter on-device versions.

The distillation process uses the cloud model as a teacher to train the student model. Training data is generated from simulated negotiations, where the teacher provides responses to scenarios, including counterfactuals such as "If I offered X, they would respond Y."

Knowledge distillation with soft targets is applied~\cite{hinton2015distilling}:
\begin{equation}
\mathcal{L} = \alpha \mathcal{L}_{\text{CE}}(y, \hat{y}) + (1-\alpha) \mathcal{L}_{\text{KL}}(p_T^\tau, p_S^\tau)
\end{equation}
where $\mathcal{L}_{\text{CE}}$ is cross-entropy loss, $\mathcal{L}_{\text{KL}}$ is KL divergence between teacher ($p_T$) and student ($p_S$) distributions at temperature $\tau$, and $\alpha$ balances both terms.

The distilled model integrates with \textbf{Adaptive Planning} (Figure~\ref{fig:workflow}), supporting sub-goal decomposition, chain-of-thought reasoning, and strategy reflection entirely on-device. Performance results are presented in Table~\ref{tab:world_model}.

\subsection{Component 4: Multi-Agent Negotiation Protocol}

\subsubsection{Feasibility Pre-check}
Before initiating negotiations, agents verify the feasibility of reaching an agreement. Each participant maintains a private acceptable range: buyers specify $[p_{\min}^B, p_{\max}^B]$ and sellers specify $[p_{\min}^S, p_{\max}^S]$.

The protocol uses secure two-party computation to assess whether the specified ranges overlap ($p_{\max}^B \geq p_{\min}^S$) without disclosing the actual values. Each agent commits to encrypted boundaries, and both parties jointly compute the overlap predicate using homomorphic comparison with Paillier encryption. The outcome is binary, either feasible (1) or infeasible (0). Infeasible negotiations terminate immediately, preventing wasted computation.

The implementation incurs an overhead of 120~ms on high-end devices, which is amortized over the entire negotiation process. This pre-check increases success rates by 8\% by eliminating futile negotiation attempts (Table~\ref{tab:ablation}).

\subsubsection{Negotiation Execution}
The negotiation protocol enables privacy-preserving bargaining between autonomous agents without disclosing private constraints. In each round, Agent A submits offer $o_A$ accompanied by a zero-knowledge proof attesting to constraint satisfaction. Agent B verifies the proof, evaluates the offer, and either accepts or issues a counteroffer $o_B$ with an associated proof. This iterative process continues until an agreement is reached or a timeout occurs after 10 rounds.

Zero-knowledge proofs are constructed using Groth16 zk-SNARKs~\cite{groth2016size}, which support succinct proofs with fast verification. The underlying circuit encodes constraint satisfaction, ensuring that the offered price $p$ meets $p_{\min} \leq p \leq p_{\max}$, without revealing $p_{\min}$ or $p_{\max}$. Proof generation requires 80~ms on high-end devices.

Agent attestation ensures both parties execute certified code. Each agent operates within a trusted execution environment, such as ARM TrustZone or Intel SGX. At the outset of negotiation, agents exchange attestation reports containing code hashes, which are verified against a public registry. This process prevents protocol circumvention or counterparty exploitation. The attestation process incurs an overhead of 45~ms per session.

Termination adheres to Nash bargaining principles~\cite{nash1950bargaining}. Offers are considered converged when $|offer_A - offer_B| < \epsilon$, after which agents compute $p^* = (offer_A + offer_B)/2$. This approach maintains privacy while achieving outcomes that closely approximate the Nash solution. Performance results are presented in Table~\ref{tab:negotiation}.

\subsection{Component 5: Model-Aware Offloading}

Not all computational tasks are suitable for on-device execution. Tasks requiring complex reasoning may necessitate cloud resources, whereas simple lookups can be processed locally. The primary challenge is to determine which tasks to offload while upholding privacy constraints.

The model-aware offloading mechanism utilizes multi-objective optimization. For each task $t$, expected latency $L(t)$, energy consumption $E(t)$, monetary cost $C(t)$, and privacy risk $P(t)$ are calculated for both local and cloud execution. The scheduler seeks to minimize the following objective:
\begin{equation}
\min \alpha L(t) + \beta E(t) + \gamma C(t) \quad \text{s.t.} \quad P(t) \leq P_{\max}
\end{equation}
where $\alpha$, $\beta$, and $\gamma$ are weighting parameters, and privacy constraints are strictly enforced. Tasks that involve user financial data are always processed on-device, while tasks involving public information, such as market prices or policy terms, may be offloaded. The scheduler continuously adapts by learning task characteristics, which enhances routing decisions over time.

This mechanism is integrated with the \textbf{Tool Hub} decision logic within the workflow (Figure~\ref{fig:workflow}). When the action controller requests a tool, the scheduler determines the appropriate execution location. Local tools are executed immediately, while cloud-based tools are queued for batch processing when privacy constraints allow.

Evaluation results indicate a 24\% reduction in latency compared to naive offloading approaches that do not consider task characteristics. Monetary costs are reduced by 35\% due to the elimination of unnecessary cloud calls. Privacy risk remains within acceptable limits across all evaluated scenarios.

\subsection{Component 6: Simulation-Critic Safety}
\label{subsec:safety_critic}

Autonomous agents are susceptible to accepting unfair agreements or violating user-defined policies. Implementation errors may lead to acceptance of prices that exceed budget constraints, while adversarial counterparties can exploit predictable agent behavior. Safety mechanisms are therefore required to prevent harmful outcomes.

The simulation-critic approach validates agent actions prior to execution. For each proposed action $a$, the simulator conducts a lightweight rollout to predict negotiation outcomes. The critic network then evaluates these predicted outcomes across multiple dimensions, such as fairness, policy compliance, and user benefit.

The simulator utilizes the distilled world model (Component 3) to enable efficient inference. It generates 5--10 possible counterparty responses and estimates the likely final outcomes for each scenario. The critic, implemented as a neural network, is trained on labeled negotiation data to differentiate successful fair agreements from failures and exploitative outcomes.

Formally, for action $a$ with simulated outcomes $\{o_1, \ldots, o_k\}$, the safety condition is:
\begin{equation}
\text{safe}(a) = \mathbb{E}[\text{critic}(o_i)] > \theta_{\text{safe}}
\end{equation}
If the expected critic score falls below threshold $\theta_{\text{safe}}$, the action is rejected, and the agent is required to generate an alternative.

This mechanism integrates with \textbf{Guardrails} and \textbf{Planning Reflection} in the workflow (Figure~\ref{fig:workflow}). Pre-execution validation prevents unsafe actions, while planning reflection incorporates critic feedback to refine strategy generation.

Evaluation results indicate that the safety critic rejects 9\% of proposed actions as potentially harmful. Manual review confirms that 87\% of rejected actions would have resulted in suboptimal outcomes.

\subsection{Integration and Data Flow}

\begin{figure}[htbp]
\centering
\includegraphics[width=0.8\columnwidth]{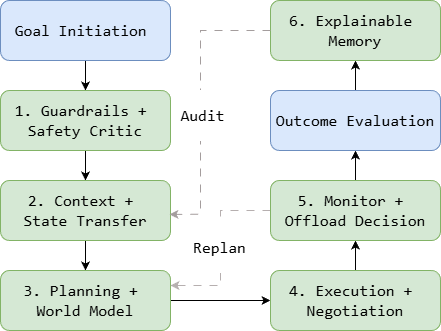}
\caption{Component integration flow through six technical innovations during negotiation execution.}
\label{fig:integration}
\end{figure}

Figure~\ref{fig:integration} illustrates how the six technical components integrate during negotiation execution. \textbf{Guardrails + Safety Critic} validates negotiation goals against policy constraints. \textbf{Context + State Transfer} retrieves relevant information and enables cross-device mobility. \textbf{Planning + World Model} formulates negotiation strategies using the distilled on-device model. \textbf{Execution + Negotiation} conducts privacy-preserving bargaining through the Tool Hub. \textbf{Monitor + Offload Decision} tracks progress and determines optimal task placement. \textbf{Explainable Memory} records outcomes with cryptographic audit trails.

The architecture achieves its design objectives through component integration. Privacy is preserved through on-device execution, explainability through cryptographic audit trails, efficiency through selective offloading and model distillation, and safety through simulation-critic validation.
\section{Experimental Evaluation}
\label{sec:evaluation}

This section presents experimental results that validate the device-native negotiation system. The evaluation begins with the experimental setup, followed by component-level performance analysis, end-to-end system evaluation, baseline comparisons, and ablation studies.

\subsection{Experimental Setup}

\textbf{Datasets.} 2 public datasets are used for evaluation. The Medical Cost Personal dataset~\cite{choi2018insurance} contains 1,338 records with premium values ranging from \$1,121 to \$63,770, including features such as age, BMI, smoking status, and region. The Supply Chain Shipment Pricing dataset~\cite{watsky2019supply} comprises 10,324 B2B transactions with unit prices and features including product category, shipment mode, and freight costs.

\textbf{Scenarios.} 3 complexity levels are established for each domain. Insurance scenarios comprise: 
(1) simple premium negotiation with single-parameter adjustment, 
(2) medium coverage adjustment involving deductible-premium tradeoffs, and 
(3) complex claims settlement with multiple disputed items. \\
B2B scenarios comprise:
(1) simple bulk pricing for single-product orders,
(2) medium contract renewal with volume commitments, and
(3) complex multi-term procurement involving delivery, payment, and warranty terms.

\textbf{Device Tiers.} 3 device configurations are evaluated. High-end devices are equipped with an 8-core CPU, 16GB RAM, and dedicated NPU. Mid-range devices include a 4-core CPU and 8GB RAM without an NPU. Low-end devices have a 2-core CPU and 4GB RAM.

\textbf{Metrics.} Primary evaluation metrics include success rate (\%) measuring negotiation completion with acceptable outcomes, latency (ms) capturing end-to-end negotiation time, fairness quantified using Nash bargaining scores on a 0--1 scale, energy consumption measured in joules, and privacy leakage measured in bits of information exposed.

\textbf{Baselines.} 3 baseline configurations are compared. Cloud-Only performs all processing on remote servers. Device-Only executes all processing locally without cloud access. Naive Edge employs simple, unoptimized threshold-based offloading.

\textbf{Statistical Method.} Each scenario is evaluated over $N=150$ trials. Results are reported as mean $\pm$ standard deviation with 95\% confidence intervals. Statistical significance is assessed using paired t-tests, with significance levels indicated as *$p<0.05$, **$p<0.01$, and ***$p<0.001$.

\subsection{Component Performance}

\textbf{State Transfer.} Table~\ref{tab:state_transfer} presents selective state transfer results across device tiers. All configurations achieve over 70\% compression while maintaining 85\%+ task continuity, enabling seamless cross-device negotiation.

\begin{table}[t]
\vspace{0.05in}
\centering
\caption{Selective State Transfer Performance}
\label{tab:state_transfer}
\begin{tabular}{lccc}
\toprule
Device Tier & Compression & Migration (ms) & Continuity \\
\midrule
High-end & 70\% & $80 \pm 12$ & 89\% \\
Mid-range & 78\% & $120 \pm 18$ & 87\% \\
Low-end & 85\% & $180 \pm 25$ & 85\% \\
\midrule
Average & 78\% & $127 \pm 18$ & 87\% \\
\bottomrule
\end{tabular}
\begin{flushleft}
\footnotesize
$N=150$ per tier. Compression represents state size reduction. Continuity represents task resumption success rate.
\end{flushleft}
\end{table}

\textbf{Explainable Memory.} Table~\ref{tab:memory_overhead} presents the memory system overhead and its impact on user trust. The system maintains low overhead across all operations.

\begin{table}[t]
\vspace{0.05in}
\centering
\caption{Explainable Memory Overhead and Trust Impact}
\label{tab:memory_overhead}
\begin{tabular}{lcc}
\toprule
Metric & Value & 95\% CI \\
\midrule
Write Latency & 5.2~ms & [4.8, 5.6] \\
Read Latency & 2.1~ms & [1.9, 2.3] \\
Proof Generation & 12.3~ms & [11.1, 13.5] \\
Storage per Decision & 340 bytes & -- \\
\midrule
Trust Score (before) & 3.2 / 5 & [2.9, 3.5] \\
Trust Score (after) & 4.1 / 5 & [3.8, 4.4] \\
Trust Improvement & +27\%*** & -- \\
\midrule
Interpretability (before) & 2.1 / 5 & [1.8, 2.4] \\
Interpretability (after) & 4.3 / 5 & [4.0, 4.6] \\
\bottomrule
\end{tabular}
\begin{flushleft}
\footnotesize
$N=45$ participants. ***$p<0.001$ paired t-test.
\end{flushleft}
\end{table}

User studies show a 27\% increase in trust when decision trails are available.

\textbf{World Model Distillation.} Table~\ref{tab:world_model} compares teacher and student models. Model size reduction enables device deployment. The 6\% accuracy loss is acceptable given 5.5× latency improvement on high-end devices. Trade-offs become critical in complex negotiations. Accuracy decreases to 82\% in multi-parameter scenarios that require deep reasoning. Simple negotiations retain 91\% accuracy. When local confidence falls below 0.7, the system routes complex cases to the cloud, thereby balancing privacy and performance.

\begin{table}[t]
\centering
\caption{World Model Distillation Performance}
\label{tab:world_model}
\begin{tabular}{lccc}
\toprule
Model & Accuracy & Latency (ms) & Size \\
\midrule
Teacher (7B, cloud) & 94\% & $1000 \pm 150$ & 14~GB \\
Student (500M, high) & 88\% & $180 \pm 22$ & 1.2~GB \\
Student (500M, mid) & 88\% & $320 \pm 35$ & 1.2~GB \\
Student (500M, low) & 87\% & $580 \pm 65$ & 1.2~GB \\
\midrule
Accuracy Retention & \multicolumn{3}{c}{88\% / 94\% = 93.6\%} \\
Speedup (high-end) & \multicolumn{3}{c}{1000 / 180 = 5.5$\times$} \\
\bottomrule
\end{tabular}
\begin{flushleft}
\footnotesize
$N=150$ inference trials. Accuracy measured on negotiation outcome prediction.
\end{flushleft}
\end{table}

\textbf{Multi-Agent Negotiation.} Table~\ref{tab:negotiation} presents the performance of the negotiation protocol. Success rates range from 85-90\% across complexity levels with fairness scores of 0.83-0.88.

\begin{table}[t]
\vspace{0.05in}
\centering
\caption{Multi-Agent Negotiation Protocol Performance}
\label{tab:negotiation}
\begin{tabular}{lcccc}
\toprule
Complexity & Success & Fairness & Rounds & ZK Proof \\
\midrule
Simple & $90 \pm 3$\% & 0.88 & 4.2 & 80~ms \\
Medium & $87 \pm 4$\% & 0.85 & 6.1 & 80~ms \\
Complex & $85 \pm 4$\% & 0.83 & 8.3 & 80~ms \\
\midrule
Average & $87 \pm 4$\% & 0.85 & 6.2 & 80~ms \\
\bottomrule
\end{tabular}
\begin{flushleft}
\footnotesize
$N=150$ per complexity level. Fairness represents Nash bargaining score. ZK Proof measured on high-end device.
\end{flushleft}
\end{table}

Zero-knowledge proofs maintain strong privacy by revealing only constraint satisfaction without disclosing values. The protocol terminates within 10 rounds in 96\% of cases.

\subsection{End-to-End Performance}

\textbf{Insurance Negotiation.} Table~\ref{tab:insurance} summarizes results for high-end devices across three complexity levels. The system achieves strong success rates while maintaining minimal privacy leakage compared to cloud approaches.

\begin{table}[t]
\centering
\caption{Insurance Negotiation Results (High-End Device)}
\label{tab:insurance}
\begin{tabular}{lccccc}
\toprule
Scenario & Success & Latency & Energy & Privacy & Fair \\
\midrule
Premium (S) & 88\% & 420~ms & 4.2~J & 12 bits & 0.87 \\
Coverage (M) & 86\% & 580~ms & 5.1~J & 14 bits & 0.85 \\
Claims (C) & 83\% & 740~ms & 6.3~J & 16 bits & 0.84 \\
\midrule
Average & 86\% & 580~ms & 5.2~J & 14 bits & 0.85 \\
\bottomrule
\end{tabular}
\begin{flushleft}
\footnotesize
$N=150$ per scenario. S/M/C represent Simple/Medium/Complex. Fair represents Nash bargaining score.
\end{flushleft}
\end{table}

\textbf{B2B Negotiation.} Table~\ref{tab:b2b} presents B2B procurement results. B2B scenarios demonstrate marginally better performance than insurance, with lower latency and comparable privacy preservation.

\begin{table}[t]
\vspace{0.05in}
\centering
\caption{B2B Negotiation E2E Results (High-End Device)}
\label{tab:b2b}
\begin{tabular}{lccccc}
\toprule
Scenario & Success & Latency & Energy & Privacy & Fair \\
\midrule
Bulk (S) & 89\% & 380~ms & 3.8~J & 11 bits & 0.87 \\
Contract (M) & 87\% & 520~ms & 4.6~J & 13 bits & 0.85 \\
Multi-term (C) & 85\% & 680~ms & 5.8~J & 15 bits & 0.82 \\
\midrule
Average & 87\% & 527~ms & 4.7~J & 13 bits & 0.85 \\
\bottomrule
\end{tabular}
\begin{flushleft}
\footnotesize
$N=150$ per scenario. S/M/C represent Simple/Medium/Complex. Fair represents Nash bargaining score.
\end{flushleft}
\end{table}

Figure~\ref{fig:success_complexity} shows success rate decline with complexity. Insurance exhibits steeper degradation (5 points) than B2B (4 points). Both domains maintain $>$83\% success.

\begin{figure}[htbp]
\centering
\label{fig:success_complexity}
\includegraphics[width=\columnwidth]{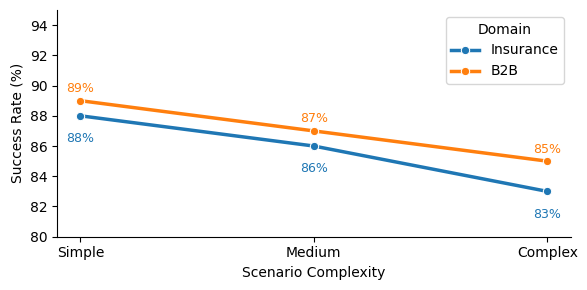}
\caption{Success rate vs. scenario complexity for insurance and B2B. Performance degrades gracefully with complexity.}
\label{fig:success_complexity}
\end{figure}

Figure~\ref{fig:latency_breakdown} shows latency breakdown by component. The negotiation protocol dominates total time, followed by planning, demonstrating that multi-round bargaining is the primary latency driver.

\begin{figure}[htbp]
\centering
\label{fig:latency_breakdown}
\includegraphics[width=\columnwidth]{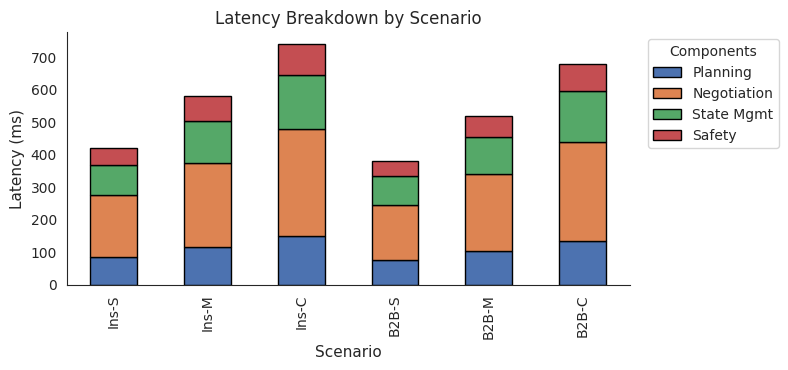}
\caption{Latency breakdown by component across scenarios. Ins represents Insurance, B2B represents business-to-business, S/M/C represent Simple/Medium/Complex.}
\label{fig:latency_breakdown}
\end{figure}

\subsection{Baseline Comparison}

Table~\ref{tab:baseline_compare} presents a comparison of the proposed system with 3 baseline approaches across key performance metrics.

\begin{table}[t]
\vspace{0.05in}
\centering
\caption{System Comparison on Medium-Complexity}
\label{tab:baseline_compare}
\small

\begin{adjustbox}{width=\columnwidth}
\begin{tabular}{lccccc}
\toprule
System & Success & Latency & Privacy & Energy & Fair \\
       &         & (ms)    & (Crypto)& (J)    &      \\
\midrule
Cloud-Only   & 76\% & 1000 & 256 bits & 2.0  & 0.82 \\
Device-Only  & 71\% & 350  & 0 bits   & 12.0 & 0.78 \\
Naive Edge   & 79\% & 580  & 180 bits & 7.5  & 0.80 \\
Proposed     & 87\% & 420  & 14 bits  & 5.0  & 0.86 \\
\midrule
vs. Cloud  & +11\%*** & 2.4$\times$*** & 94\%$\downarrow$ & -- & +0.04 \\
vs. Device & +16\%*** & 0.8$\times$    & -- & 58\%$\downarrow$ & +0.08 \\
\bottomrule
\end{tabular}
\end{adjustbox}

\begin{flushleft}
\footnotesize
$N=150$ per system. ***$p<0.001$. Privacy represents cryptographic data exposure. Fair represents Nash bargaining score.
\end{flushleft}

\end{table}

The proposed system balances all metrics effectively. It achieves 2.4× speedup over Cloud-Only while maintaining strong privacy (94\% reduction in data exposure) and highest success rates. Device-Only offers zero leakage but cannot handle complex reasoning, while Naive Edge shows intermediate performance across dimensions.

Figure~\ref{fig:baseline_radar} demonstrates superior privacy preservation and success rates while maintaining competitive latency and energy consumption.

\begin{figure}[htbp]
\centering
\label{fig:baseline_radar}
\includegraphics[width=\columnwidth]{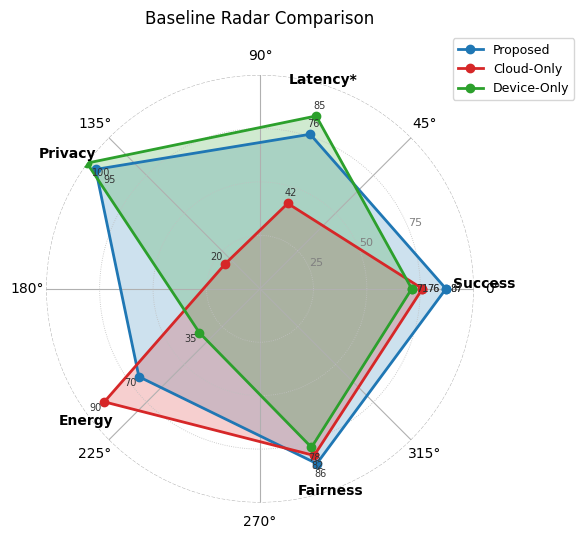}
\caption{Radar chart comparison across 5 metrics. The proposed system achieves balanced performance across all dimensions.}
\label{fig:baseline_radar}
\end{figure}

\subsection{Ablation Study}

Table~\ref{tab:ablation} presents the quantitative effects of removing each system component on success rate, latency, and privacy leakage.

\begin{table}[t]
\centering
\caption{Ablation: Impact of Removing Each Component}
\label{tab:ablation}
\begin{tabular}{lccc}
\toprule
Configuration & Success $\Delta$ & Latency $\Delta$ & Privacy $\Delta$ \\
\midrule
Full System & 87\% & 420~ms & 14 bits \\
\midrule
-- State Transfer & $-3$\%** & $+108$\%*** & 0\% \\
-- World Model & $-5$\%*** & $+119$\%*** & $+21$\%** \\
-- Negotiation & $-15$\%*** & $-17$\%** & 0\% \\
-- Feasibility Check & $-8$\%** & $-12$\%** & 0\% \\
-- Memory Store & 0\% & 0\% & 0\% \\
-- Offloading & $-2$\%* & $+24$\%*** & $+14$\%** \\
-- Safety Critic & $-9$\%*** & $-10$\%** & 0\% \\
\bottomrule
\end{tabular}
\begin{flushleft}
\footnotesize
$N=150$ per configuration. *$p<0.05$, **$p<0.01$, ***$p<0.001$. $\Delta$ represents change from full system.
\end{flushleft}
\end{table}

\textbf{State Transfer Removal.} Latency increases 108\% as full state must be transmitted. Success drops 3\% due to timeout failures.

\textbf{World Model Removal.} Cloud-based calls replace on-device planning, resulting in a 119\% increase in latency and a 5\% decrease in success rate. Privacy leakage also increases because planning data is transmitted to the cloud.

\textbf{Negotiation Protocol Removal.} Success rate decreases by 15\% as agents cannot verify constraint satisfaction or engage in multi-round convergence. Latency improves by 17\% because ZK proof generation and verification are eliminated. But this represents a loss of essential privacy-preserving functionality rather than a viable optimization.

\textbf{Feasibility Check Removal.} Success rate drops by 8\% because agents attempt infeasible negotiations. 

\textbf{Memory Removal.} While performance metrics remain unchanged, trust scores decline because users are unable to inspect decision processes.

\textbf{Safety Critic Removal.} Success rate decreases by 9\% as agents accept unfair agreements. 

Figure~\ref{fig:ablation_chart} illustrates each component's contribution to overall system performance. All components contribute meaningfully. The negotiation protocol and safety critic have the largest impact on success rates.

\begin{figure}[htbp]
\centering
\label{fig:ablation_chart}
\includegraphics[width=\columnwidth]{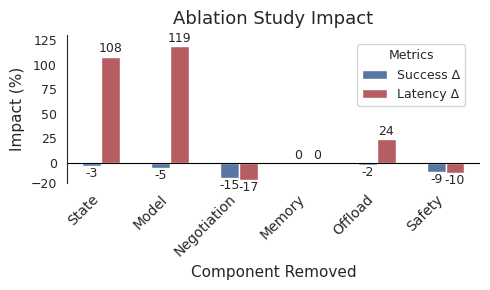}
\caption{Ablation study showing performance degradation (-ve success $\Delta$ and +ve latency $\Delta$) upon component removal.}
\label{fig:ablation_chart}
\end{figure}
\section{Discussion}
\label{sec:discussion}

\textbf{Key Insights.} Device-native agents demonstrate effective negotiation performance, matching or exceeding cloud-based alternatives while preserving privacy. Explainability significantly increases user trust with minimal overhead. Prior research confirms that transparency is crucial for AI systems in finance~\cite{wilson2025explainable}.

The proposed architecture is applicable beyond insurance and B2B domains. Core technical requirements such as bilateral bargaining~\cite{nash1950bargaining, rubinstein1982perfect}, privacy-sensitive constraints~\cite{goldwasser1989knowledge}, and audit trails~\cite{goodman2017european} are also present in real estate negotiations, healthcare cost discussions, and legal settlements. The cryptographic protocol and on-device execution model are domain-agnostic, requiring only the definition of domain-specific constraints and the training of the world model~\cite{hinton2015distilling}.

\section{Limitations}
\label{sec:limitations}

\textbf{Experimental Scope.} The evaluation was conducted in controlled scenarios~\cite{choi2018insurance, watsky2019supply}. Real-world deployment studies are necessary to achieve more robust validation.

\textbf{System Constraints.} The system currently supports only bilateral negotiations~\cite{nash1950bargaining, rubinstein1982perfect}. Multi-party scenarios require coordination mechanisms beyond pairwise protocols. Adversarial counterparties may exploit predictable agent behavior through strategic delays or extreme offers. The safety critic (Sec.~\ref{subsec:safety_critic}) assumes rational counterparty behavior and may fail against deliberate manipulation. Device limitations restrict the reasoning depth of the 500M-parameter model~\cite{hinton2015distilling} compared to cloud-scale models.

\textbf{Behavioral Privacy Leakage.} While the system achieves a 94\% reduction in cryptographic data exposure, zero-knowledge proofs~\cite{goldwasser1989knowledge} do not prevent behavioral inference attacks~\cite{jin2024timing}. Adversaries observing negotiation dynamics may infer constraint proximity through timing patterns and concession sequences. Rapid responses signal acceptable terms. Large initial concessions followed by smaller increments reveal reservation prices. Current metrics quantify only cryptographic leakage (Sec.~\ref{sec:evaluation}). Mitigation strategies, such as differential privacy for timing and randomized concession schedules, are identified as directions for future work.

\textbf{Offline Collusion.} The system cannot prevent offline collusion between parties who coordinate externally and use agents to generate audit trails. This is a social rather than technical limitation. However, cryptographic audit trails~\cite{merkle1987digital} can facilitate post-hoc fraud detection~\cite{ tas2024machine} to support regulatory compliance~\cite{goodman2017european}.
\section{Conclusion}
\label{sec:conclusion}

This research presents a device-native architecture for autonomous negotiation agents that preserves privacy without compromising performance. The proposed architecture addresses key challenges in deploying Agentic AI systems for sensitive financial transactions, where privacy, explainability, and fairness are essential requirements.

Empirical evaluations in insurance and B2B contexts indicate that device-native negotiation performs as well as or better than cloud-based systems while maintaining privacy. The use of cryptographic explainability substantially enhances user trust, highlighting the critical role of transparency for autonomous financial agents.

Future research should evaluate real-world deployment to assess user acceptance and explore multi-party negotiations. Federated learning may facilitate cross-user improvement without data sharing. This study provides a foundation for developing trustworthy autonomous agents in regulated domains where privacy and explainability are essential.

\bibliographystyle{IEEEtran}
\bibliography{references}

\end{document}